\documentclass[a4paper,11pt]{article}
\usepackage{caption}
\usepackage{subcaption}
\usepackage{epsfig, graphicx, subfloat, float}
\usepackage{latexsym}
\usepackage{cite}
\usepackage{amssymb}
\usepackage{latexsym}
\usepackage{mathrsfs}
\usepackage{amsmath}
\usepackage{epstopdf}
\usepackage{authblk}
\usepackage{color}
\usepackage[usenames]{xcolor}
\graphicspath{Image/}
\begin{document}
\title {Gravitational Baryogenesis in Rainbow Gauss-Bonnet gravity }
\author[1]{P. Goodarzi \footnote{parviz.goodarzi@abru.ac.ir}}
\author[2]{ H. Mohseni Sadjadi\footnote{mohsenisad@ut.ac.ir}}
\author[2]{R. Darabi \footnote{rdarabi@ut.ac.ir}}
\affil[1]{Department of Physics, Faculty of Basic Sciences, Ayatollah Boroujerdi University, Boroujerd, Iran}
\affil[2]{ Department of Physics, University of Tehran, Tehran, Iran}

\maketitle
\begin{abstract}
The gravitational baryogenesis is studied in regularized four dimensional Einstein Gauss Bonnet model with rainbow gravity corrections. This combined framework generates a non vanishing time derivative of the Ricci scalar even during radiation domination era. We derive analytical expressions for the baryon to entropy ratio in both the high curvature and perturbative regimes. The Gauss Bonnet contribution produces nonzero asymmetry precisely at radiation era, where the Einstein Hilbert contribution vanishes. Numerical analysis identifies viable parameter spaces reproducing the observed baryon to entropy ratio
compatible with observations.

\end{abstract}

\section{Introduction}
One of the fundamental questions in modern cosmology is the origin of the matter-antimatter asymmetry observed in the Universe. Cosmological observations, including those from the Cosmic Microwave Background (CMB) and Big Bang Nucleosynthesis (BBN) indicate a baryon-to-entropy ratio of order $Y_B = \frac{n_B}{s} \sim 10^{-10}$ \cite{Planck2018,PDG2022} (where $n_B$ denotes the net baryon number density and $s$ is the entropy density), a small but nonzero value
that requires a dynamical explanation beyond the Standard Model of particle physics.

Since Sakharov seminal work \cite{Sakharov1967}, three conditions have been recognized as necessary for successful baryogenesis:  baryon number violation, violation of $C$ and $CP$ symmetries, and departure from thermal equilibrium. While numerous mechanisms have been proposed, including GUT baryogenesis \cite{KolbTurner,RiottoTrodden}, Affleck-Dine scenarios\cite{AffleckDine}, electroweak baryogenesis \cite{KuzminRubakovShaposhnikov}, and leptogenesis \cite{FukugitaYanagida}, they typically attribute the asymmetry to microscopic processes in the matter sector, with gravity playing only an indirect role through cosmic expansion.
Spontaneous baryogenesis provides another possibility, allowing asymmetry generation even in thermal equilibrium through effective CPT-violating interactions \cite{CohenKaplan},\cite{spon0,spon1}.

A paradigm shift was introduced by \cite{Davoudiasl2004} through the concept of gravitational baryogenesis, wherein spacetime curvature directly contributes to baryon asymmetry generation via an effective coupling between the baryon current
and the derivative of the Ricci scalar\cite{gb1,gb2,gb3,gb4}. This induces a time-dependent, $CPT$-violating chemical potential for the baryon number. This mechanism enables the generation of baryon asymmetry even in thermal equilibrium, provided that baryon-number-violating processes remain active. However, the original proposal faces two significant challenges: the appearance of higher-order derivatives leads to Ostrogradsky's ghost instabilities \cite{DeFelice2010}, moreover during the radiation-dominated era, $R=0$ (where $R$ is the Ricci scalar) renders the mechanism ineffective precisely when baryogenesis is expected to occur.

Several modifications have addressed these issues through various approaches: extending to
modified gravity \cite{Lambiase2006,sad1,par1,mod1,mod2,mod3,mod4,mod5}; considering a time dependent equation of state parameter \cite{sad2},
employing the Gauss-Bonnet invariant
$\mathcal{G}$ instead of $R$ \cite{Odintsov2016}, introducing nonlinear curvature couplings \cite{Feng2008,par2}, incorporating bulk viscosity effects \cite{Zimdahl1996,Harko2011}, or exploring gravitational leptogenesis \cite{LambiaseMohanty2007,FujiiYanagida2002}.

Among these, the Gauss-Bonnet extension is particularly attractive as it naturally arises in Lovelock gravity and string-inspired effective theories \cite{Lovelock1971,Zwiebach1985,BoulwareDeser1985}, yet in four dimensions it becomes a topological invariant and does not contribute dynamically.

The recent regularization scheme proposed by Glavan and Lin \cite{GlavanLin2020} and subsequently refined through scalar-tensor formulations \cite{GB1,GB2,Gurses2020,LuPangPope2020,Kobayashi2020,Aoki2020} has revived interest in four-dimensional Einstein-Gauss-Bonnet (EGB) gravity by taking the
$D\to 4$ limit of the D-dimensional action while introducing an auxiliary scalar field via conformal transformation. The resulting theory belongs to the Horndeski class, ensuring second-order field equations and freedom from Ostrogradsky instabilities. This regularized $4D$ EGB gravity provides a natural arena for exploring high-curvature modifications to Einstein gravity.

Simultaneously, Rainbow Gravity \cite{MagueijoSmolin} has emerged as a promising phenomenological approach to quantum gravity, positing an energy-dependent spacetime metric \cite{NojiriOdintsov,ata}. In this context the  metric of Friedmann- Lemaitre-Robertson-Walker (FLRW) space time is given by
\begin{equation}
ds^{2}(E) = -\frac{dt^{2}}{f^{2}(E)} + \frac{a^{2}(t)}{g^{2}(E)} \, \delta_{ij}\, dx^{i} dx^{j},
\end{equation}
that encodes modified dispersion relations at high energies. $a(t)$ is the cosmological scale factor that describes the expansion of the universe, and $f(E)$ and $g(E)$ are the rainbow functions that introduce energy-dependent modifications into the structure of spacetime. These functions are defined in such a way that they approach unity in the low-energy limit, ensuring that the standard FLRW metric and the usual results of general relativity are recovered in this regime \cite{MagueijoSmolin}.
Given that the background metric in this framework is energy-dependent, the derived field equations inherently incorporate a dependence on the rainbow functions and, consequently, on the energy scale. This implies that the cosmological evolution within this model is shaped by a dual set of influences: firstly, the higher-order curvature corrections stemming from the Gauss–Bonnet term, and secondly, the altered geometric structure of spacetime due to the energy-dependent metric characteristic of rainbow gravity \cite{MagueijoSmolin,Zwiebach}. The interplay of these two effects potentially offers a robust framework for examining the universe's behavior at extremely high energies.

Despite the individual successes of these frameworks, a comprehensive investigation of gravitational baryogenesis in the combined setting of regularized $4D$ EGB gravity with Rainbow Gravity corrections has remained unexplored. Such a synthesis is particularly motivated by the observation that both frameworks modify the high-energy regime of cosmological evolution, the Gauss-Bonnet term through higher-curvature corrections and Rainbow Gravity through energy-dependent geometry—potentially offering novel solutions to the radiation-era problem of gravitational baryogenesis.

The paper is organized as follows: In Section 2, we present the theoretical framework of regularized 4D EGB gravity in Rainbow Gravity, deriving the modified Friedmann equations and scalar field dynamics. Section 3 analyzes gravitational baryogenesis in the GB dominated regime, while Section 4 addresses the perturbative regime. Section 5 presents comprehensive numerical analyses identifying viable parameter spaces. We conclude in Section 6 with a summary of our findings and discussion of future directions.

We use units $\hbar=c=1$
\section{Regularized Four-Dimensional Einstein--Gauss--Bonnet in Rainbow Gravity}

To construct the regularized four-dimensional Einstein--Gauss--Bonnet (EGB) theory, we begin with the $D$-dimensional action introduced in Refs.~\cite{GB1,GB2}
\begin{eqnarray}\label{actionD}
S_{GB}&=&\frac{1}{16\pi G_D}\left[\int d^Dx\,\sqrt{-g}\,R+\frac{\alpha}{D-4}\left(\int d^Dx\,\sqrt{-\tilde{g}}\,\tilde{\mathcal{G}}-\int d^Dx\,\sqrt{-g}\,\mathcal{G}\right)\right]\nonumber \\
&+&S_m .
\end{eqnarray}
where $G_D$ denotes the gravitational constant in $D$ spacetime dimensions and $\alpha$ is the GB coupling parameter. The term $S_m$ represents the matter action.

In this construction, an auxiliary metric $\tilde{g}_{\mu\nu}$ is introduced, which is related to the spacetime metric $g_{\mu\nu}$ through the conformal transformation
\begin{equation}\label{conformal}
\tilde{g}_{\mu\nu}=e^{\phi} g_{\mu\nu},
\end{equation}
where $\phi$ is a scalar field that appears as a consequence of the regularization procedure. The quantity $\tilde{\mathcal{G}}$ denotes the GB invariant constructed from the metric $\tilde{g}_{\mu\nu}$, while $\mathcal{G}$ is the standard GB invariant defined by
\begin{equation}\label{GB}
\mathcal{G}=R^2-4R_{\mu\nu}R^{\mu\nu}+R_{\mu\nu\rho\sigma}R^{\mu\nu\rho\sigma}.
\end{equation}

The prescription above leads to a well-defined and finite limit when $D\rightarrow4$. Performing this limit, the theory reduces to an effective four dimensional scalar--tensor model described by the action
\begin{eqnarray}\label{action4}
&&S_{GB}=\frac{1}{16\pi G}\int d^4x\,\sqrt{-g}\Big[
R+\alpha\Big(
\phi \mathcal{G}
+4G^{\mu\nu}\nabla_\mu\phi\nabla_\nu\phi
-4\nabla_\alpha\nabla^\alpha\Box\phi
\nonumber \\
&&+2(\nabla_\mu\phi\nabla^\mu\phi)^2
\Big)
\Big]
+S_m .
\end{eqnarray}

The resulting theory belongs to the Horndeski class of scalar-tensor theories, which constitutes the most general class of scalar--tensor models yielding second-order equations of motion. Consequently, the theory is free from the Ostrogradsky instability typically associated with higher-derivative gravitational theories.

Varying the action with respect to the metric $g_{\mu\nu}$ leads to the modified gravitational field equations
\begin{equation}\label{metricfield}
\begin{aligned}
\frac{1}{M_P^2}T_{\mu\nu} =\;& G_{\mu\nu}
+\alpha\Big\{
-2R\!\left[(\nabla_\mu\phi)(\nabla_\nu\phi)+\nabla_\mu\nabla_\nu\phi\right]  \\
&+8R^\beta_{(\mu}\!\left[\nabla_{\nu)}\nabla_\beta\phi
+(\nabla_{\nu)}\phi)(\nabla_\beta\phi)\right]
-2G_{\mu\nu}\!\left[(\nabla\phi)^2+2\Box\phi\right] \\
&-4\!\left[(\nabla_\mu\phi)(\nabla_\nu\phi)+\nabla_\mu\nabla_\nu\phi\right]\Box\phi
-\!\left[g_{\mu\nu}(\nabla\phi)^2-4(\nabla_\mu\phi)(\nabla_\nu\phi)\right](\nabla\phi)^2 \\
&+8(\nabla_{(\mu}\phi)(\nabla_{\nu)}\nabla_\beta\phi)\nabla^\beta\phi
-4g_{\mu\nu}R^{\beta\sigma}\!\left[\nabla_\beta\nabla_\sigma\phi
+(\nabla_\beta\phi)(\nabla_\sigma\phi)\right] \\
&+2g_{\mu\nu}(\Box\phi)^2
-2g_{\mu\nu}(\nabla_\beta\nabla_\sigma\phi)(\nabla^\beta\nabla^\sigma\phi)
+4g_{\mu\nu}(\nabla^\beta\phi)(\nabla^\sigma\phi)(\nabla_\beta\nabla_\sigma\phi) \\
&+4(\nabla_\beta\nabla_\nu\phi)(\nabla^\beta\nabla_\mu\phi)
+4R_{\mu\beta\nu\sigma}\!\left[(\nabla^\beta\phi)(\nabla^\sigma\phi)
+\nabla^\sigma\nabla^\beta\phi\right]
\Big\}.
\end{aligned}
\end{equation}
where $M_P$ denotes the reduced Planck mass and $T_{\mu\nu}$ is the energy--momentum tensor of the matter fields.
In addition, variation of the action with respect to the scalar field $\phi$ yields the corresponding scalar field equation
\begin{eqnarray}\label{scalarfield}
8\nabla_\nu\nabla_\mu\phi\nabla^\nu\nabla^\mu\phi
+16\nabla^\mu\phi\nabla^\nu\phi\nabla_\nu\phi\nabla_\mu\phi
-8(\Box\phi)^2\\
+8(\nabla\phi)^2\Box\phi
+8R^{\mu\nu}\nabla_\mu\phi\nabla_\nu\phi\nonumber
+8G^{\mu\nu}\nabla_\nu\nabla_\mu\phi
=\mathcal{G}.
\end{eqnarray}

Equations~(\ref{metricfield}) and (\ref{scalarfield}) describe the coupled dynamics of the spacetime geometry and the scalar degree of freedom emerging from the regularized GB sector.
To investigate the cosmological implications of the regularized four-dimensional EGB gravity, we consider a homogeneous and isotropic FLRW universe in the context of Rainbow Gravity. The modified spacetime metric is parameterized as~\cite{MagueijoSmolin,RG2}
\begin{equation}\label{5}
ds^2 = -\frac{dt^2}{f^2(t)} + \frac{a^2(t)}{g^2(t)} \left( dr^2 + r^2 d\Omega^2 \right),
\end{equation}
where $f(t)$ and $g(t)$ are the Rainbow functions that incorporate quantum-gravity corrections in the high-energy limit, $a(t)$ is the cosmic scale factor, and $d\Omega^2 = d\theta^2 + \sin^2\theta d\phi^2$ denotes the metric of a unit 2-sphere. By defining the cosmic rainbow time coordinate $\tau$ and the effective scale factor $A(\tau)$ via the relations
\begin{equation}\label{time_redef}
d\tau = \frac{dt}{f(t)}, \quad \text{and} \quad A(\tau) = \frac{a(t)}{g(t)},
\end{equation}
we rewrite the metric (\ref{5}) as
\begin{equation}\label{51}
ds^2 = -d\tau^2 + A^2(\tau) \left( dr^2 + r^2 d\Omega^2 \right).
\end{equation}
This representation allows us to describe the cosmic evolution using the effective Hubble parameter defined with respect to the rainbow time, $\mathcal{H} \equiv A'/A$, where a prime denotes a derivative with respect to $\tau$ ($d/d\tau$).

In the framework of Rainbow Gravity, the modification of the space--time geometry is described through rainbow functions that depend on the energy scale of the system \cite{MagueijoSmolin,Ling2007}. To investigate the inflationary dynamics, we parameterize the rainbow function using a power-law form \cite{Awad2013,Ali2014}

\begin{equation}
f(t)^{2} = 1 + \left(\frac{H}{M}\right)^{2\beta},
\end{equation}
where $H$ denotes the Hubble parameter, $M$ represents a characteristic energy scale (typically of the order of the Planck scale), and $\beta>0$ is the rainbow parameter that characterizes the strength of the quantum-gravity corrections. In the late-time universe, where $H/M \ll 1$, the second term becomes negligible and the rainbow function reduces to $f(t) \approx 1$, recovering the standard cosmological behavior. However, during the inflationary epoch the condition $H^{2} \gg M^{2}$ holds, and therefore the rainbow function can be approximated as
\begin{equation}
f(t) \approx \left(\frac{H}{M}\right)^{\beta}.
\label{f(t)}
\end{equation}
We assume that the Universe is dominated by perfect fluids, for which the energy-momentum tensor is given by
\begin{equation}\label{stress_tensor}
T_{\mu\nu} = (\rho + P) u_\mu u_\nu + P g_{\mu\nu},
\end{equation}
where $\rho$ and $P$ represent the energy density and pressure, respectively. For a comoving observer with the four-velocity $u^\mu = (1, 0, 0, 0)$, the energy-momentum tensor takes the diagonal form $T^\mu_\nu = \text{diag}(-\rho, P, P, P)$.
Substituting the metric~(\ref{51}) into the modified Einstein field equations~(\ref{metricfield}), the $0-0$ component yields the first modified Friedmann equation in the form
\begin{equation}\label{6}
\alpha \left( -\phi'^4 + 4\phi'^3 \mathcal{H} - 6\phi'^2 \mathcal{H}^2 + 4\phi' \mathcal{H}^3 \right) + \mathcal{H}^2 = \frac{1}{3M_P^2} \rho.
\end{equation}
Simultaneously, the equation of motion for the scalar field $\phi$ is derived by evaluating Eq.~(\ref{scalarfield}) in the FLRW background, which yields
\begin{equation}\label{7}
\left( A \phi'' + A'\phi' - A'' \right) \left( A^2\phi'^2 - 2AA'\phi' + A'^2 \right) = 0.
\end{equation}
Equation~(\ref{7}) can be integrated directly to obtain the first-order differential equation for the scalar field
\begin{equation}\label{8}
\phi' = \mathcal{H} + \frac{C}{A},
\end{equation}
where $C$ is an arbitrary integration constant.

By substituting the scalar field solution~(\ref{8}) back into the modified Friedmann equation~(\ref{6}), the dynamics of the scale factor $A(\tau)$ is governed by the simplified relation
\begin{equation}\label{9}
\alpha \mathcal{H}^4 + \mathcal{H}^2 = \frac{1}{3M_P^2} \rho + \alpha \frac{C^4}{A^4}.
\end{equation}
To determine the acceleration equation, we evaluate the spatial components of the modified field equations~(\ref{metricfield}), yielding the second modified Friedmann equation
\begin{equation}\label{10}
3\alpha \mathcal{H}^4 + 4\alpha \mathcal{H}^2 \mathcal{H}' + 3\mathcal{H}^2 + 2\mathcal{H}' = -\frac{P}{M_P^2} - \frac{\alpha C^4}{A^4}.
\end{equation}
Finally, by eliminating the $\mathcal{H}^4$ term between Eqs.~(\ref{9}) and (\ref{10}), we arrive at the dynamical equation for the Hubble parameter
\begin{equation}\label{11}
\mathcal{H}' \left( 1 + 2\alpha \mathcal{H}^2 \right) = -\frac{1}{2M_P^2} (P + \rho) - \frac{2\alpha C^4}{A^4}.
\end{equation}
Equations~(\ref{9}) and (\ref{11}) govern the background cosmological evolution of the regularized EGB gravity in the presence of Rainbow Gravity corrections.

It is worth noting that if we choose a non-vanishing integration constant, $C \neq 0$, an effective dark radiation component emerges, which can be defined as $\rho_r \equiv 3M_P^2 \alpha \frac{C^4}{A^4}$. In the limit where $f = g = 1$ and $C = 0$, one recovers the standard cosmological equations of the regularized four-dimensional EGB theory~\cite{GlavanLin2020,LuPangPope}.

We now turn to the generation of the baryon asymmetry of the Universe within the framework of regularized four-dimensional EGB gravity supplemented by Gravity's Rainbow. We employ the standard gravitational baryogenesis mechanism, in which the derivative of the Ricci scalar is coupled to the baryon current $J^\mu$ through the effective interaction~\cite{Davoudiasl2004,LambiaseMohanty2007}
\begin{equation}
S_{\rm int} = \frac{1}{M_*^2} \int d^4x \sqrt{-g}\, (\partial_\mu R) J^\mu,
\end{equation}
where $M_*$ denotes the cutoff scale of the effective theory.

In a homogeneous and isotropic cosmological background, only the temporal component of the derivative contributes to the interaction. As a result, this coupling induces an effective chemical potential for baryons and antibaryons
\begin{equation}
\mu_B = \frac{\dot{R}}{M_*^2},
\end{equation}
where the dot denotes differentiation with respect to the cosmic rainbow time. In the high-temperature regime, where $|\mu_B| \ll T$, the net baryon number density can be approximated as
\begin{equation}
n_B \simeq -\frac{g_b}{6}\mu_B T^2,
\end{equation}
with $g_b$ denoting the intrinsic degrees of freedom of baryons. Substituting the above expression for $\mu_B$, one obtains
\begin{equation}
n_B \simeq -\frac{g_b}{6}\frac{\dot{R}}{M_*^2}T^2.
\end{equation}

Using the standard entropy density of a relativistic thermal bath~\cite{KolbTurner}
\begin{equation}
s = \frac{2\pi^2}{45} g_* T^3,
\end{equation}
where $g_*$ is the effective number of relativistic degrees of freedom, the baryon-to-entropy ratio at the decoupling temperature $T_D$ is given by~\cite{Davoudiasl2004,LambiaseMohanty2007}
\begin{equation}
Y_B \equiv \frac{n_B}{s}
\simeq
-\frac{15 g_b}{4\pi^2 g_*}\frac{\dot{R}}{M_*^2 T}\bigg|_{T=T_D}.
\label{YBbasic}
\end{equation}
This expression will be used in the following to constrain the model parameters in light of the observed baryon asymmetry.

We now compute $\dot R$ in the Rainbow--EGB cosmological background.
We ignore the dark radiation, $ C=0$ \cite{GB2}. With this choice, Eq.~\eqref{9} reduces to
\begin{equation}
    \alpha\mathcal{H} ^4 + \mathcal{H}^2
    =
    \frac{\rho}{3M_P^2}.
    \label{eq:modified_friedmann_Czero}
\end{equation}

 The Ricci scalar for the metric (\ref{51}) is given by
\begin{equation}\label{12}
R=6\mathcal{H}'+12\mathcal{H}^2=6f^2(t)\left(\dot{H}-\dot{\tilde{G}}+2(H-\tilde{G})^2\right)+6f(t)\dot{f(t)}(H-G)
\end{equation}
where $\tilde{G}=\frac{\dot{g}}{g}$.
In most studies of Rainbow Gravity, simplifying assumptions are often adopted for the rainbow functions. In this work, we restrict our analysis to the case in which $g(t) = 1$, so that all modifications of the space--time geometry are encoded solely in the function $f(t)$. In the following, we derive the corresponding solutions for a power-law scale factor within this scenario.

To determine the resulting baryon asymmetry in this framework, we investigate two limiting regimes of the parameter $\alpha \mathcal{H}^2$, namely $\alpha \mathcal{H}^2 \gg 1$ and $\alpha \mathcal{H}^2 \ll 1$. These limits correspond to different dynamical regimes of the model and enable a systematic analysis of the baryogenesis mechanism.

\section{ regime  $\alpha \mathcal{H}^2 \gg 1$}
At sufficiently high energy,  the quartic term dominates and one obtains
\begin{equation}
\alpha \mathcal H^4
\simeq
\frac{\rho}{3M_P^2}.
\label{HEfriedmann}
\end{equation}

Using Friedman equations (\ref{9}) and (\ref{11}), we obtain
\begin{equation}\label{13}
R=\frac{\sqrt{3}}{\sqrt{\alpha M_P^2}}\left(-\frac{3}{2}(w+1)+4\right)\sqrt{\rho}
\end{equation}
We have considered a constant equation of state parameter, $w$, $P=w \rho$.
From the continuity equation
\begin{equation}\label{14}
\rho'+3\mathcal{H}(w+1)\rho=0
\end{equation}
we obtain
\begin{equation}
\dot{R}= \frac{1}{f(t)}\left(-\frac{3^{\frac{5}{4}}}{2(\alpha M_P^2)^{\frac{3}{4}}}\left(4-\frac{3}{2}(w+1)\right)\left(w+1\right)\rho^{\frac{3}{4}}\right).
\label{Rdot}
\end{equation}

Here \(\rho\) should be regarded as the effective background energy density
entering the modified Friedmann equation. To express the result in terms of
the temperature, we use the standard relativistic thermal-bath
estimate
\begin{equation}
\rho
=
\frac{\pi^2}{30}g_*T^4 .
\label{rho_T}
\end{equation}
In the analysis to follow we shall assume that a thermal equilibrium exists, so in all cases which we study, we will
assume that the Universe evolves slowly from an equilibrium state to an equilibrium state, with the energy density
being related to the temperature $T$ as Eq.(\ref{rho_T}).
Substituting Eq.(\ref{rho_T}) into Eq.~\eqref{Rdot}, we find
\begin{equation}
\dot R
=
-\frac{3^{5/4}}{4(\alpha M_P^2)^{3/4}}
(5-3w)(1+w)
\left(
\frac{\pi^2g_*}{30}
\right)^{3/4}
\frac{T^3}{f(T)} .
\label{RdotT}
\end{equation}

Using  Eq.~\eqref{f(t)} and since the physical Hubble parameter is related to \(\mathcal H\) by \(H=\mathcal H/f(T)\), we obtain
\begin{equation}
\frac{1}{f(T)}
=
\left(
\frac{M}{\mathcal H}
\right)^{
\frac{\beta}{1+\beta}
}.
\label{f_T}
\end{equation}

From Eq.~\eqref{HEfriedmann} and the radiation energy density, the Hubble parameter in the high-energy EGB regime is
\begin{equation}
\mathcal H
=
\left(
\frac{\pi^2g_*}{90\alpha M_P^2}
\right)^{1/4}T .
\label{HT}
\end{equation}
Substituting Eq.~\eqref{HT} into Eq.~\eqref{f_T}, one finds
\begin{equation}
\dot R
=
-\frac{9}{4}
(5-3w)(1+w)
M^{\frac{\beta}{1+\beta}}
\left(
\frac{\pi^2g_*}{90\alpha M_P^2}
\right)^{
\frac{3+2\beta}{4(1+\beta)}
}
T^{
\frac{3+2\beta}{1+\beta}
}.
\label{Rdotfinal}
\end{equation}

Substituting Eq.~\eqref{Rdotfinal} into the baryon-to-entropy ratio in Eq.~\eqref{YBbasic}, evaluated at \(T=T_D\), gives

\begin{equation}
Y_B
=
\frac{135g_b}{16\pi^2g_*}
(5-3w)(1+w)
\left(
\frac{\pi^2g_*}{90\alpha M_P^2}
\right)^{
\frac{3+2\beta}{4(1+\beta)}
}
\frac{
M^{\frac{\beta}{1+\beta}}
T_D^{\frac{2+\beta}{1+\beta}}
}{M_*^2}.
\label{YBfinal}
\end{equation}

This expression shows that the baryon asymmetry depends explicitly on the GB coupling \(\alpha\), the rainbow scale \(M\), the rainbow parameter \(\beta\), and the decoupling temperature \(T_D\). In contrast with the standard Ricci-scalar gravitational baryogenesis scenario in ordinary radiation-dominated Einstein cosmology, the present high-energy EGB regime gives a non-vanishing \(\dot R\) even for the radiation equation of state. For \(w=1/3\), one obtains
\begin{equation}
Y_B(w=1/3)
=
\frac{45g_b}{\pi^2g_*}
\left(
\frac{\pi^2g_*}{90\alpha M_P^2}
\right)^{
\frac{3+2\beta}{4(1+\beta)}
}
\frac{
M^{\frac{\beta}{1+\beta}}
T_D^{\frac{2+\beta}{1+\beta}}
}{M_*^2}.
\end{equation}

Hence, in the Rainbow--EGB cosmological framework, Ricci-scalar gravitational baryogenesis can generate a nonzero baryon asymmetry during the radiation-dominated epoch.

Next, we discuss the origin of the $B$-violating interaction that is required in any baryogenesis scenario.
To keep the discussion model–independent, we assume that the baryon number violating processes are induced
by an effective operator $\mathcal{O}_B$ of mass dimension $4+n$ ,
where $n > 0$ measures how far the operator is from being renormalizable.
Such an operator generates baryon-number violating interactions suppressed by a mass scale $M_B$, so that
the corresponding interaction rate scales as
\begin{equation}
\Gamma_{B} \sim \frac{T^{2n+1}}{M_B^{\,2n}} .
\label{Gamma}
\end{equation}

At very high temperatures, the rate $\Gamma_B$ is larger than the Hubble expansion rate $H$, and thus
baryon-number violating interactions remain in thermal equilibrium. As the Universe expands and cools down,
the temperature drops and the interaction rate decreases faster than the Hubble rate. Eventually, at some
temperature $T \simeq T_D$, the rate falls below the Hubble expansion rate,
\begin{equation}
\Gamma_{B}(T_D) \simeq H(T_D),
\label{Gamma_H}
\end{equation}
marking the decoupling of $B$-violating processes. For temperatures $T < T_D$, the reaction rate becomes
too small to maintain thermal equilibrium, and baryon-number violating interactions effectively freeze out.

This decoupling temperature $T_D$ plays a central role in gravitational baryogenesis, since the baryon
asymmetry produced by the CP-violating interaction is evaluated at the moment when baryon-number
violating processes go out of equilibrium.

Using Eqs.~(\ref{f_T}) and (\ref{HT}), together with the relation $\mathcal{H}=H f(T)$, the Hubble parameter can be written as a function of the temperature.
\begin{equation}
H(T)=
\left(
\frac{\pi^2 g_*}{90\,\alpha M_P^2}
\right)^{\frac{1}{4(1+\beta)}}
M^{\frac{\beta}{1+\beta}}
T^{\frac{1}{1+\beta}} .
\label{H_T}
\end{equation}
By equating Eqs.~(\ref{H_T}) and (\ref{Gamma}), the decoupling temperature $T_D$ can be obtained in terms of the model parameters.
\begin{equation}
T_D=
\left[
M_B^{2n}
\left(
\frac{\pi^2 g_*}{90\,\alpha M_P^2}
\right)^{\frac{1}{4(1+\beta)}}
M^{\frac{\beta}{1+\beta}}
\right]^{
\frac{1}{2n+1-\frac{1}{1+\beta}}
}.
\end{equation}
Substituting the decoupling temperature $T_D$ into Eq.~(\ref{YBfinal}), the baryon asymmetry can be written entirely in terms of the model parameters as
\begin{eqnarray}
Y_B &=& \frac{135 g_b}{16 \pi^2 g_*} (5-3w)(1+w) M_*^{-2} M_B^{\frac{2n(2+\beta)}{\beta(2n+1)+2n}} \nonumber \\
&&\times M^{\frac{2n\beta+3\beta+2\beta^2}{(1+\beta)(\beta(2n+1)+2n)}} \left( \frac{\pi^2 g_*}{90 \alpha M_P^2} \right)^{\frac{4n\beta^2+2\beta^2+10n\beta+4\beta+6n+2}{4(1+\beta)(\beta(2n+1)+2n)}}
\end{eqnarray}

\section{ regime  $\alpha \mathcal{H}^2 \ll 1$}

Equation~\eqref{eq:modified_friedmann_Czero} admits two branches for $\mathcal{H}^2$, namely
\begin{equation}
\mathcal{H}^2
=
\frac{-1 \pm \sqrt{1+\dfrac{4\alpha\rho}{3M_P^2}}}{2\alpha}.
\label{eq:H2_two_branches}
\end{equation}
In the regime where the GB correction is subdominant,  (equivalently $|\alpha|\rho/M_P^2\ll 1$), one may expand the square root in Eq.~\eqref{eq:H2_two_branches}. Choosing the branch that continuously reproduces standard GR as $\alpha\to 0$, we obtain
\begin{equation}
\mathcal{H}^2
\simeq
\frac{\rho}{3M_P^2}
-
\frac{\alpha\rho^2}{9M_P^4}
+\mathcal{O}(\alpha^2).
\label{eq:H2_small_alpha}
\end{equation}
Taking the limit $\alpha\rightarrow 0$, Eq.~\eqref{eq:H2_small_alpha} reduces to the usual Friedmann equation  \cite{Weinberg2008},
\begin{equation}
\mathcal{H}^2=\frac{\rho}{3M_P^2}.
\label{eq1:H2_small_alpha}
\end{equation}
In this perturbative regime ,
the modified Hubble parameter can be expanded to first order in $\alpha$ as
\begin{equation}
\mathcal{H} \simeq
\frac{\rho^{1/2}}{\sqrt{3} M_P}
-
\frac{\alpha \rho^{3/2}}{6\sqrt{3} M_P^3}.
\label{mathcal_H}
\end{equation}

Differentiating with respect to conformal time $\tau$, and keeping terms up to $\mathcal{O}(\alpha)$, we obtain
\begin{equation}
\mathcal{H}' =
-\frac{1+w}{2}\frac{\rho}{M_P^2}
+
\frac{\alpha(1+w)}{3}\frac{\rho^2}{M_P^4}.
\end{equation}

To compute the baryon asymmetry, we next determine the Ricci scalar.
Substituting the modified Friedmann equation into the expression for the Ricci scalar in the rainbow background yields
\begin{equation}
R=
\frac{1-3w}{M_P^2}\rho
+
\frac{2\alpha(1+3w)}{3M_P^4}\rho^2.
\end{equation}
Using the conservation equation and expanding the resulting expression consistently to first order in the GB parameter $\alpha$, we obtain

\begin{equation}
\dot R=
-\frac{\sqrt{3}(1+w)\rho^{3/2}}{M_P^3}
\left(
\frac{\sqrt{3}M M_P}{\rho^{1/2}}
\right)^{\frac{\beta}{\beta+1}}
\left[
(1-3w)
+
\frac{\alpha\rho}{6M_P^2}
\left(
7+27w
+
\frac{\beta}{\beta+1}(1-3w)
\right)
\right].
\label{rdot3}
\end{equation}

By substituting Eq.~\eqref{mathcal_H} into Eq.~\eqref{f(t)} and expanding to first order in $\alpha$, the inverse rainbow function becomes

\begin{equation}
\frac{1}{f(\rho)}
\simeq
\left(
\frac{\sqrt{3}M M_P}{\rho^{1/2}}
\right)^{\frac{\beta}{\beta+1}}
\left[
1+
\frac{\beta}{\beta+1}
\frac{\alpha\rho}{6M_P^2}
\right].
\end{equation}

Substituting into $\dot R$ and retaining terms up to $\mathcal{O}(\alpha)$, the final result becomes
\begin{equation}
\dot R=
-\frac{\sqrt{3}(1+w)\rho^{3/2}}{M_P^3}
\left(
\frac{\sqrt{3}M M_P}{\rho^{1/2}}
\right)^{\frac{\beta}{\beta+1}}
\left[
(1-3w)
+
\frac{\alpha\rho}{6M_P^2}
\left(
7+27w
+
\frac{\beta}{\beta+1}(1-3w)
\right)
\right].
\label{rdot3}
\end{equation}

Using Eq.~\eqref{rho_T}, the expression for $\dot R$ can be written entirely in terms of the decoupling temperature $T=T_D$.
Finally,the baryon asymmetry is calculated as follows:

\begin{equation}
\begin{aligned}
Y_B \simeq&
\frac{15\sqrt{3}\,g_b(1+w)}{4\pi^2 g_*}
\frac{1}{M_*^2M_P^3}
\left(\sqrt{3}M M_P\right)^{\frac{\beta}{\beta+1}}
\left(\frac{\pi^2 g_*}{30}\right)^{\frac{2\beta+3}{2(\beta+1)}}
T_D^{\frac{3\beta+5}{\beta+1}}
\\
&\times
\left[
(1-3w)
+
\frac{\alpha\pi^2 g_*T_D^4}{180M_P^2}
\left(
7+27w
+
\frac{\beta}{\beta+1}(1-3w)
\right)
\right].
\end{aligned}
\end{equation}

Applying the freeze-out condition discussed in Eq.~\eqref{Gamma_H} we obtain
The decoupling temperature $T_D$ is determined by the freeze-out condition
. Using Eq.~\eqref{mathcal_H} in the rainbow function, we obtain
\begin{equation}
H(T)
\simeq
M^{\frac{\beta}{\beta+1}}
\left(
\frac{\pi \sqrt{g_*}}{\sqrt{90}M_P}
T^2
\right)^{\frac{1}{\beta+1}}
\left[
1-
\frac{\alpha\pi^2 g_*T^4}{180(\beta+1)M_P^2}
\right].
\end{equation}

Substituting Eq.~\eqref{Gamma} into the freeze-out condition, we obtain
\begin{equation}
M^{\frac{\beta}{\beta+1}}
\left(
\frac{\pi \sqrt{g_*}}{\sqrt{90}M_P}
\right)^{\frac{1}{\beta+1}}
T_D^{\frac{2}{\beta+1}}
\left[
1-
\frac{\alpha \pi^2 g_* T_D^4}{180(\beta+1)M_P^2}
\right]
=
\frac{T_D^{2n+1}}{M_B^{2n}} .
\end{equation}

For convenience, we introduce the parameter
\begin{equation}
p \equiv 2n + 1 - \frac{2}{\beta + 1}.
\end{equation}

The freeze-out condition can therefore be written as
\begin{equation}
T_D^p \simeq
M_B^{2n}
M^{\frac{\beta}{\beta+1}}
\left(
\frac{\pi \sqrt{g_*}}{\sqrt{90}M_P}
\right)^{\frac{1}{\beta+1}}
\left[
1-
\frac{\alpha \pi^2 g_* T_D^4}{180(\beta+1)M_P^2}
\right].
\end{equation}

This expression determines the decoupling temperature in the presence of rainbow gravity and GB corrections, explicitly showing how the parameters $\beta$ and $\alpha$ modify the freeze-out scale relative to the standard cosmological scenario $(\alpha=0,\beta=0)$~\cite{KolbTurner,Weinberg2008}.

\section{ Numerical analyses}

\begin{figure}[t]
\begin{center}
  \scalebox{0.52}{\includegraphics{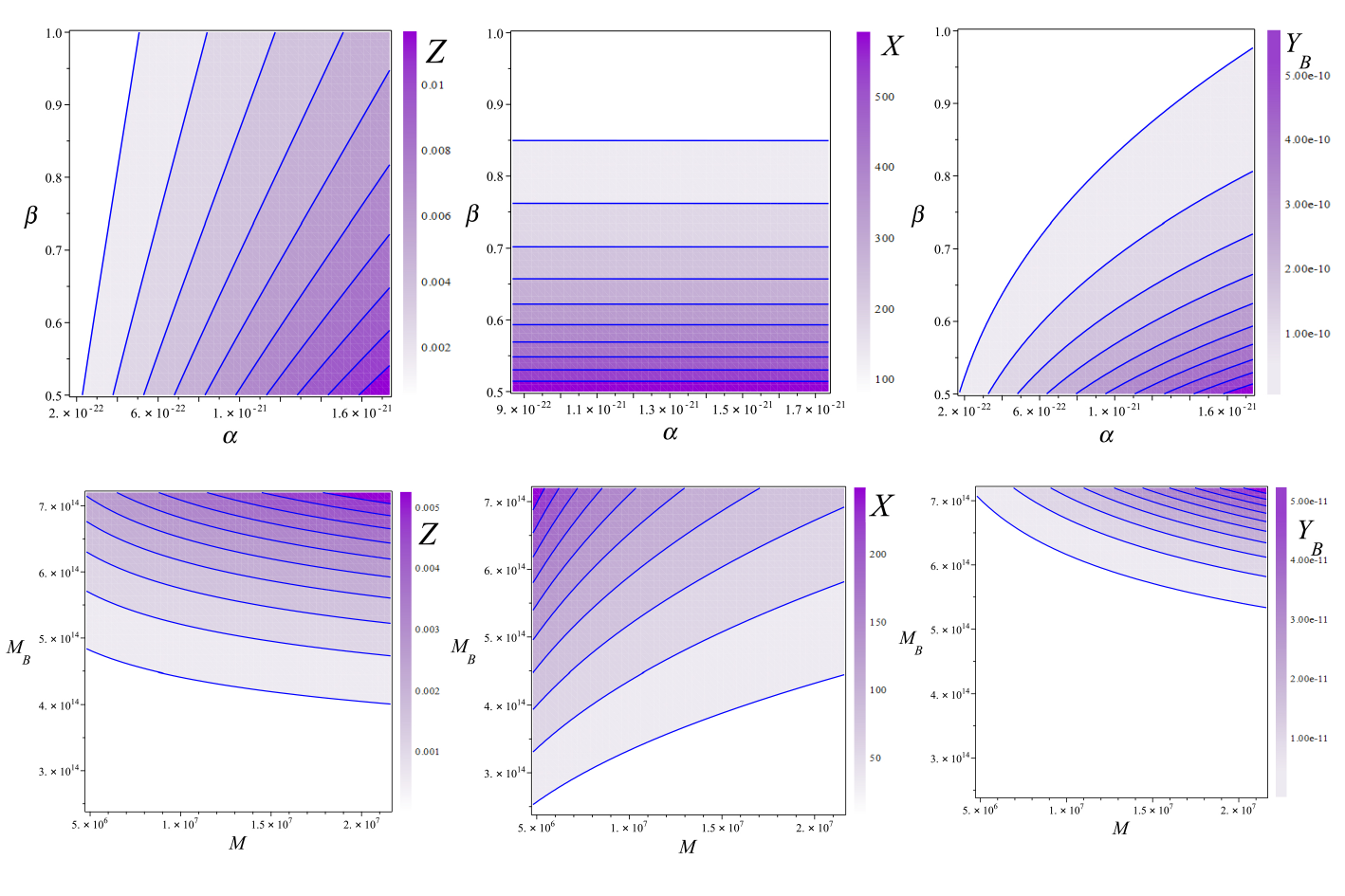}}
     \end{center}
\caption{ \footnotesize Contour plots are presented to identify plausible values for the parameters $\alpha$, $\beta$, $M$ and $M_B$, under the constraints $Z\ll1$ and $X\gg1$. The top three plots show results for  $w=1/3$, $n=2$, $M=9\times10^{-12}M_p$, $M_B=3\times10^{-4}M_p$, and $M_\star=10^{-8}M_p$. The bottom three plots illustrate results for, $\alpha=10^{16}M_p^{-2}$, $\beta=1$, $w=1/3$, $n=2$, and $M_\star=10^{-8}M_p$.}
  \label{fig1}
\end{figure}

\begin{figure}[t]
\begin{center}
  \scalebox{0.60}{\includegraphics{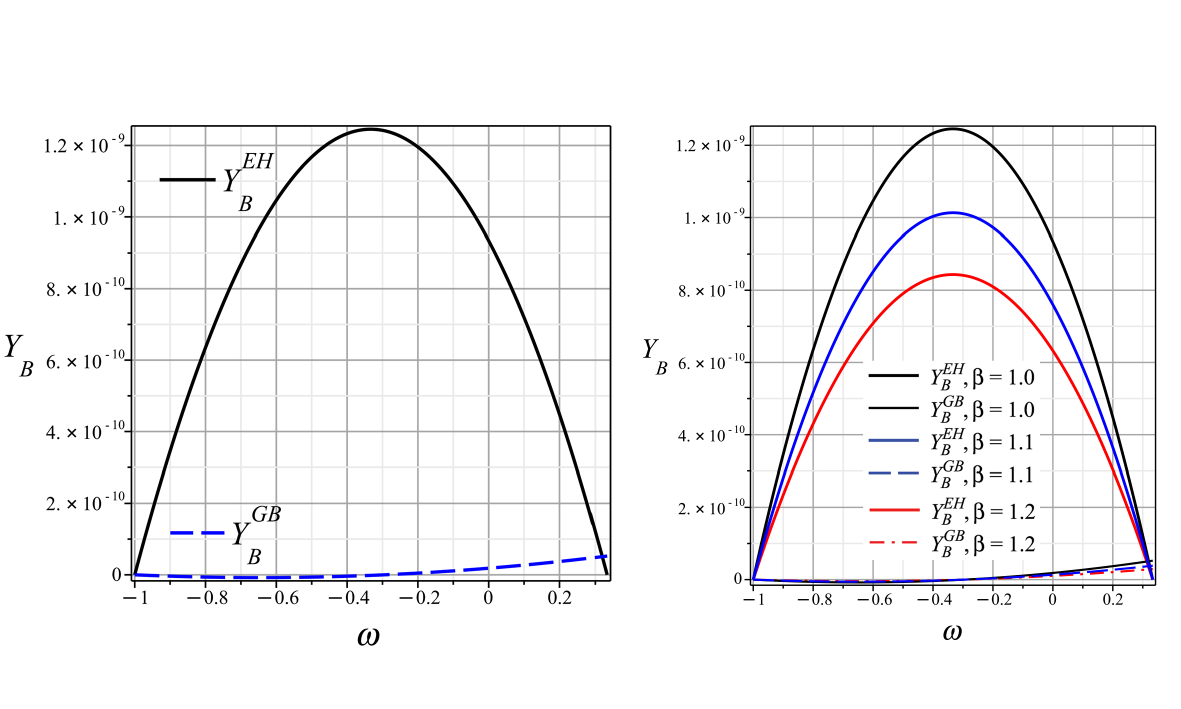}}
     \end{center}
    \caption{ \footnotesize We present a comparison of the baryon-to-entropy ratio for the Einstein-Hilbert action ($Y_B^{EH}$, solid lines) and the Gauss-Bonnet action ($Y_B^{GB}$, dashed lines) as a function of the equation of state $w$. These results are shown for several values of the rainbow parameter $\beta\in\{1.00,1.05,1.10\}$ within the regime $Z\ll1$. In these plots we assume that, $\alpha=10^{16}M_p^{-2}$, $\beta=1$, $n=2$, $M=9\times10^{-12}M_p$, $M_B=3\times10^{-4}M_p$ and $M_\star=10^{-8}M_p$.}
  \label{fig2}
\end{figure}

\begin{figure}[t]
\begin{center}
  \scalebox{0.50}{\includegraphics{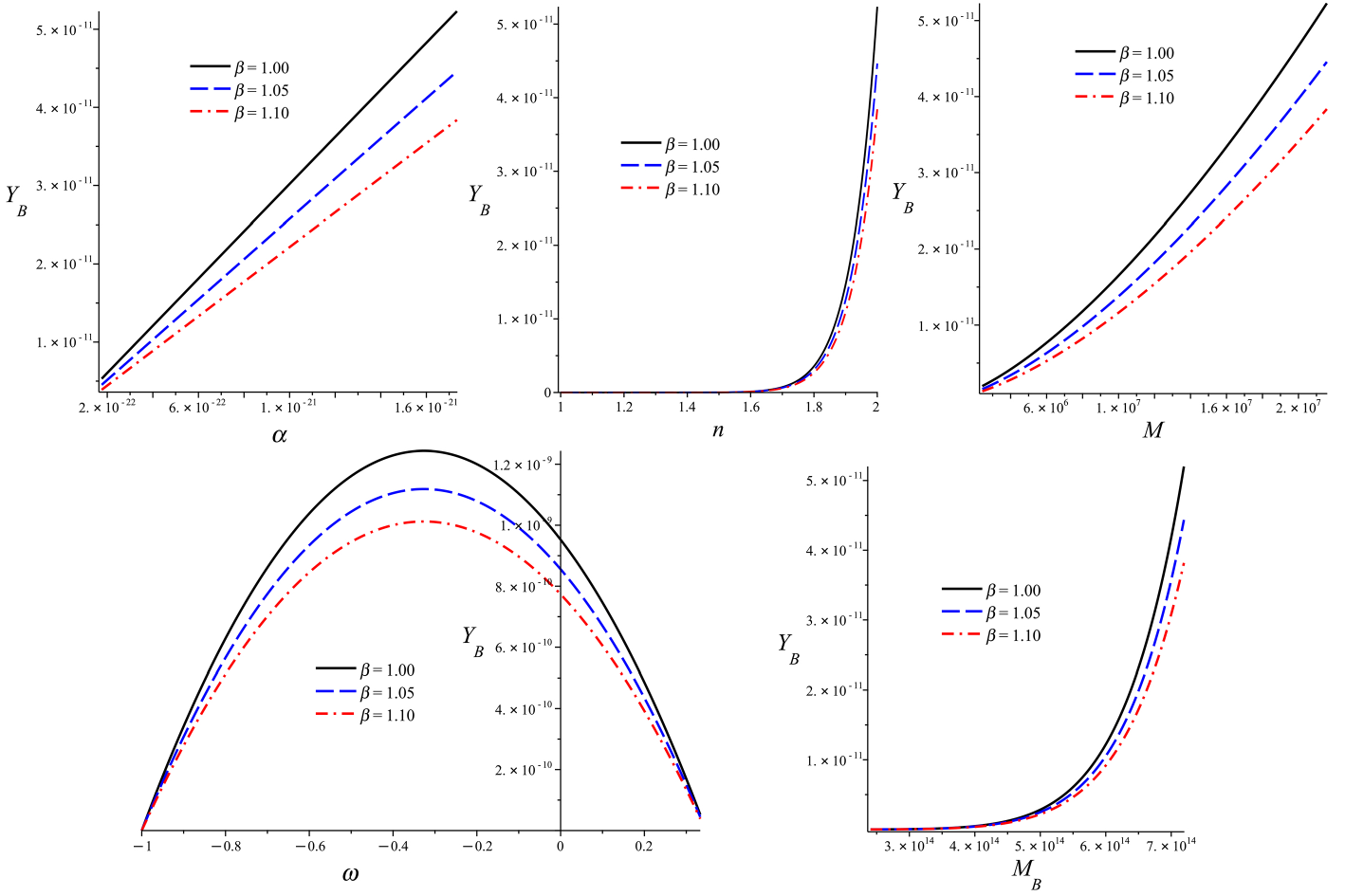}}
     \end{center}
    \caption{ \footnotesize We present the baryon-to-entropy ratio as a function of $\alpha$, $n$, $M$, $M_B$, and $w$, for several different values of the rainbow parameter $\beta\in\{1.00,1.05,1.10\}$ within the regime $Z\ll1$. For any parameters not varied in the plots, we have adopted the following values: $\alpha=10^{16}M_p^{-2}$, $\beta=1$, $n=2$, $M=9\times10^{-12}M_p$, $M_B=3\times10^{-4}M_p$ and $M_\star=10^{-8}M_p$.}
  \label{fig3}
\end{figure}

\begin{figure}[t]
\begin{center}
  \scalebox{0.50}{\includegraphics{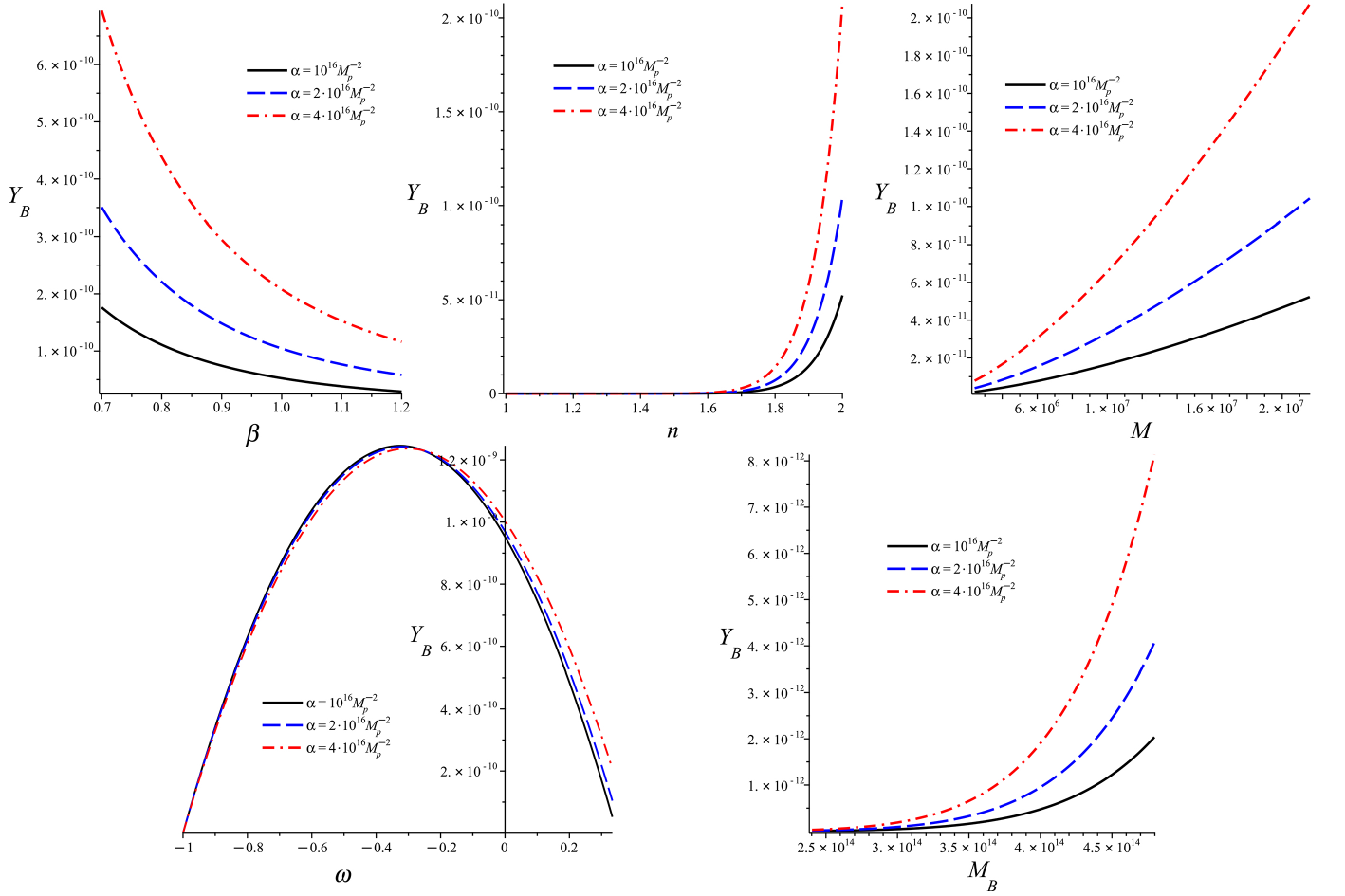}}
     \end{center}
    \caption{ \footnotesize We present the baryon-to-entropy ratio as a function of $\beta$, $n$, $M$, $M_B$, and $w$, for several different values of the Gauss-Bonnet coupling constant $\alpha\in\{10^{16}M_p^{-2},2\times10^{16}M_p^{-2},4\times10^{16}M_p^{-2}\}$ within the regime $Z\ll1$. For any parameters not varied in the plots, we have adopted the following values: $\beta=1$, $\beta=1$, $n=2$, $M=9\times10^{-12}M_p$, $M_B=3\times10^{-4}M_p$ and $M_\star=10^{-8}M_p$.}
  \label{fig4}
\end{figure}

\begin{figure}[t]
\begin{center}
  \scalebox{0.50}{\includegraphics{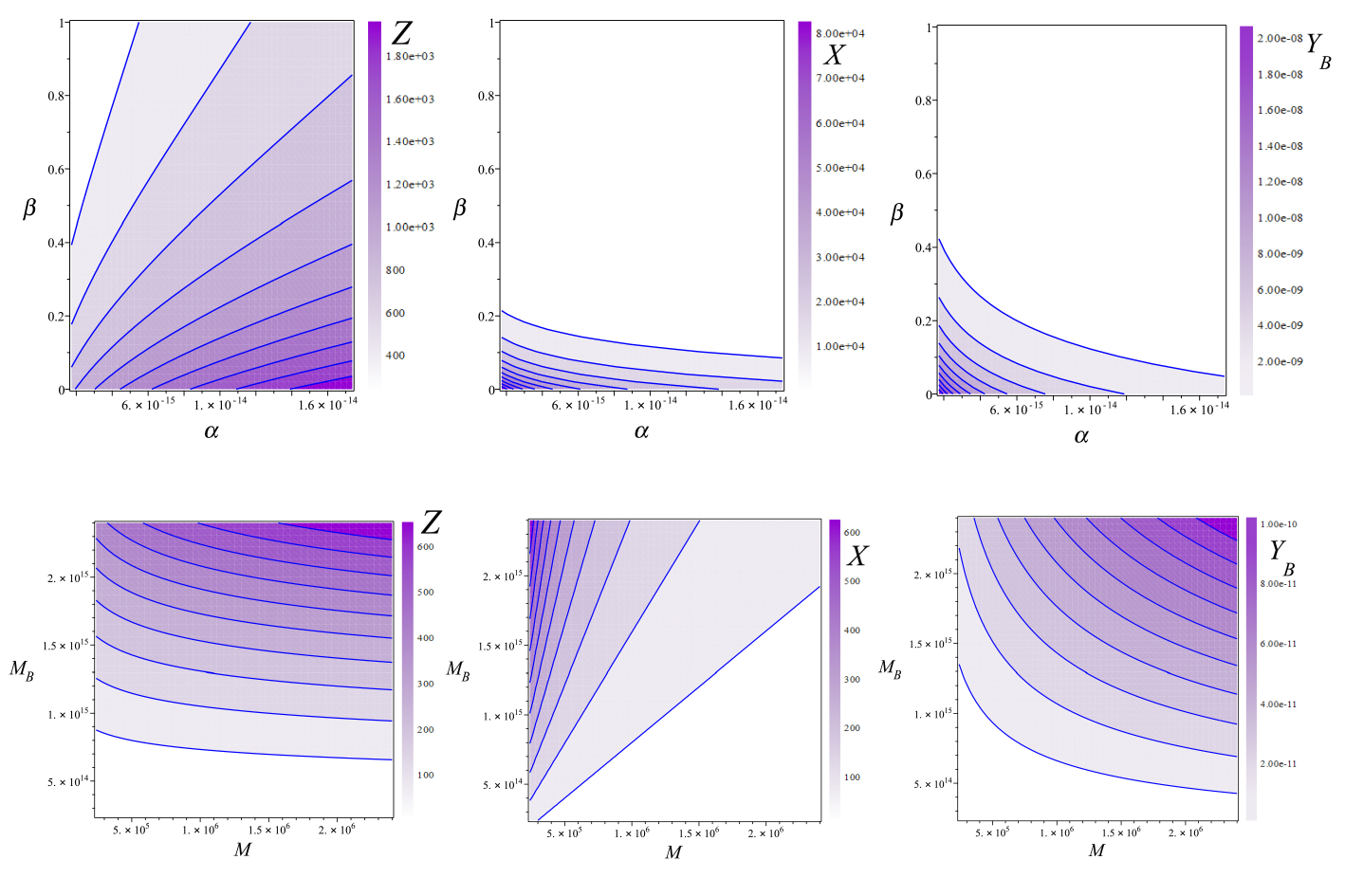}}
     \end{center}
    \caption{ \footnotesize Contour plots are presented to identify plausible values for the parameters $\alpha$, $\beta$, $M$ and $M_B$, under the constraints $Z\gg1$ and $X\gg1$. The top three plots show results for  $w=1/3$, $n=2$, $M=10^{-12}M_p$, $M_B=10^{-4}M_p$, and $M_\star=10^{-9}M_p$. The bottom three plots illustrate results for, $\alpha=10^{23}M_p^{-2}$, $\beta=1$, $w=1/3$, $n=2$, and $M_\star=10^{-9}M_p$.}
  \label{fig5}
\end{figure}

\begin{figure}[t]
\begin{center}
  \scalebox{0.50}{\includegraphics{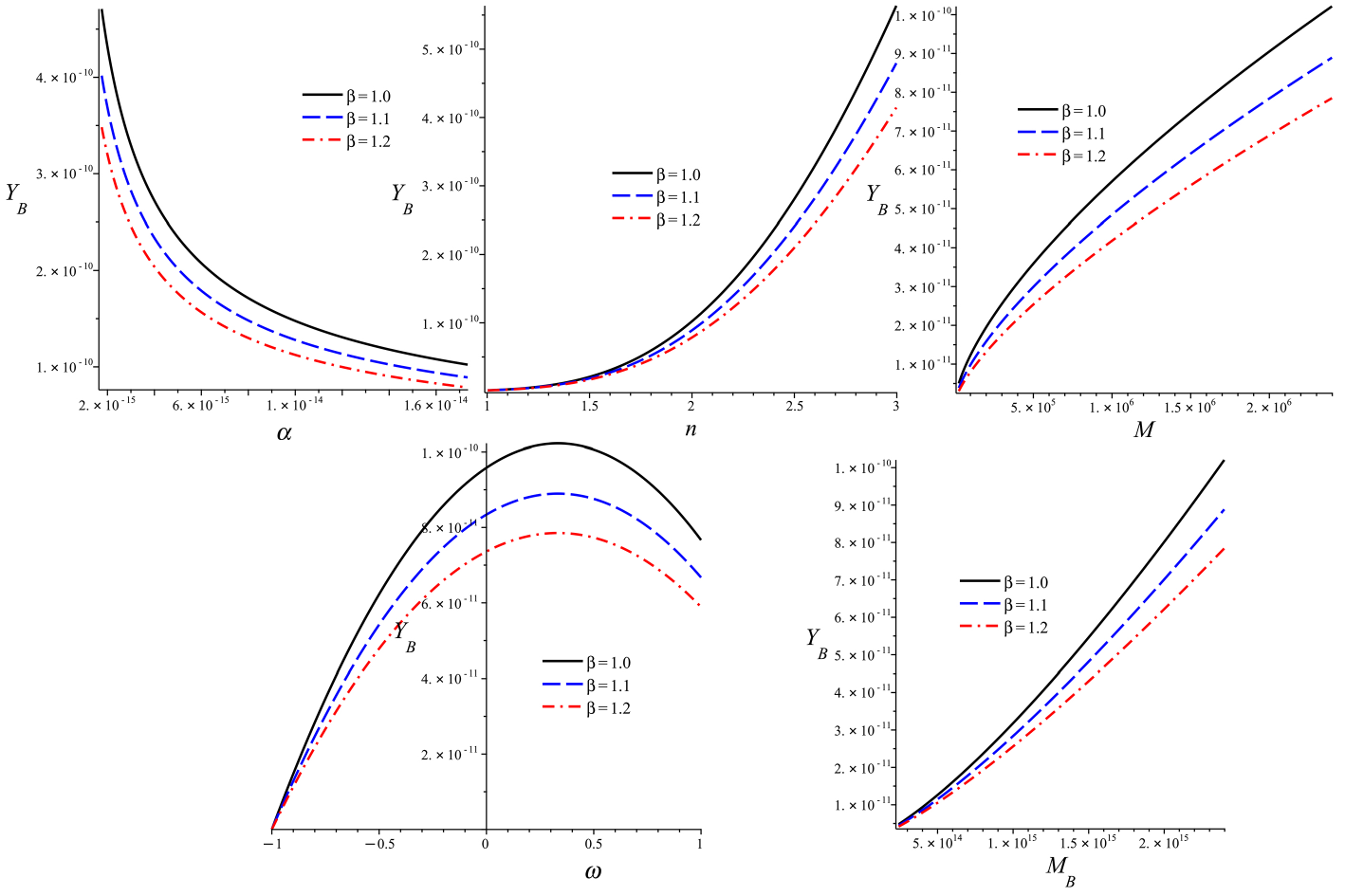}}
     \end{center}
    \caption{ \footnotesize We present the baryon-to-entropy ratio as a function of $\alpha$, $n$, $M$, $M_B$, and $w$, for several different values of the rainbow parameter $\beta\in\{1.0,1.1,1.2\}$ within the regime $Z\gg1$. For any parameters not varied in the plots, we have adopted the following values: $w=1/3$, $\alpha=10^{23}M_p^{-2}$, $\beta=1$, $n=2$, $M=10^{-12}M_p$, $M_B=10^{-3}M_p$ and $M_\star=10^{-9}M_p$.}
  \label{fig6}
\end{figure}

\begin{figure}[t]
\begin{center}
  \scalebox{0.50}{\includegraphics{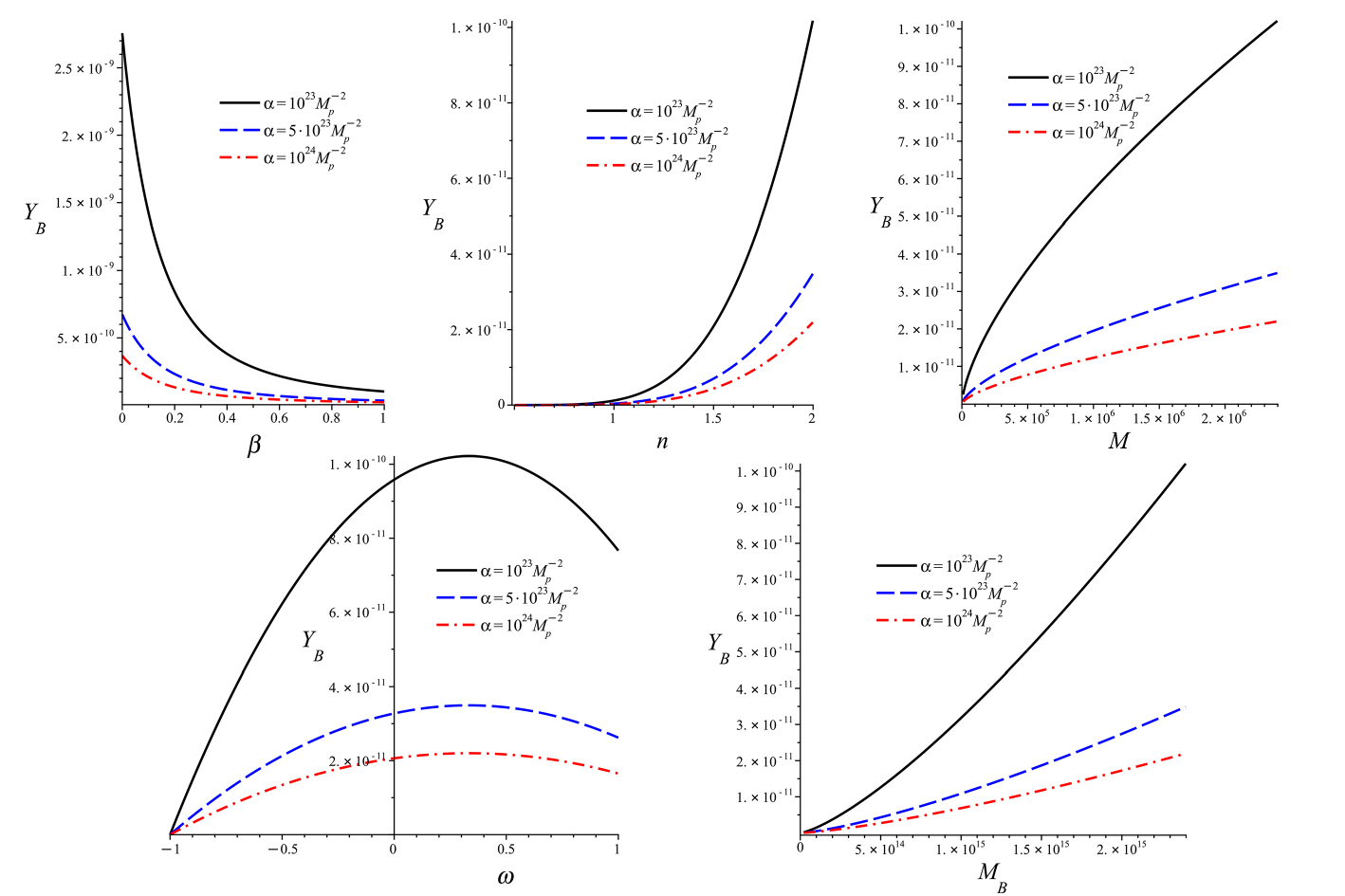}}
     \end{center}
    \caption{ \footnotesize  We present the baryon-to-entropy ratio as a function of $\beta$, $n$, $M$, $M_B$, and $w$, for several different values of the Gauss-Bonnet coupling constant $\alpha\in\{10^{23}M_p^{-2},5\times10^{23}M_p^{-2},10^{24}M_p^{-2}\}$ within the regime $Z\gg1$. For any parameters not varied in the plots, we have adopted the following values: $w=1/3$, $\beta=1$, $n=2$, $M=10^{-12}M_p$, $M_B=10^{-3}M_p$ and $M_\star=10^{-9}M_p$.}
  \label{fig7}
\end{figure}

In the preceding section, we analytically investigate gravitational baryogenesis in the context of four-dimensional EGB gravity in rainbow universe, focusing on the two limiting regimes $\alpha \mathcal{H}^2 \ll 1 $ and $\alpha \mathcal{H}^2 \gg 1$. In the following section, we extend these results through numerical analysis.

Our analysis provides the baryon asymmetry expressed as a function of the model parameters for both $\alpha \mathcal{H}^2 \ll 1 $ and $\alpha \mathcal{H}^2 \gg 1$ regimes.

We used a constant equation of state $w$, assumed that $\tilde{f}\approx (H/M)^\alpha$ during the early universe, and utilized the rate of generating B-violating interactions $\Gamma_B=T^{2n+1}/M_B^{2n}$.
We have identified seven parameters in the model, denoted as
 $ w,\alpha, \beta,n, M, M_B, M_{\star}$. These parameters are determined by requiring the resulting baryon-to-entropy ratio to be consistent with current cosmological observations while simultaneously satisfying the theoretical constraints of the model.
The model reduces to general relativity when $\alpha=0$ and $\beta=0$. Rainbow gravity corresponds to the case where $\alpha=0$ and $\beta\neq0$. When $\alpha\neq 0$ and $\beta=0$, the theory is identified as 4D GB. Finally, 4D GB in gravity’s rainbow is realized when $\alpha\neq 0$ and $\beta\neq0$.
 We define two constraints in this model with two quantity as

\begin{eqnarray}\label{30}
X\equiv{H^2\over M^2}, \\
\label{30.1}
Z\equiv\alpha \mathcal{H}^2.
\end{eqnarray}

In models where the universe is dominated by a perfect fluid with an equation of state parameter $w=1/3$ (the radiation-dominated epoch), $\dot{R}$ typically vanishes, precluding the generation of a net baryon asymmetry. Consequently, we focus our investigation on this era to analyze the specific dynamics of gravitational baryogenesis within a radiation-dominated background.

\subsection{regime $\alpha\mathcal{H}^2\ll 1$}

While the GB term was previously treated merely as a perturbative correction in this regime, we now explore its capacity to generate baryon asymmetry in a manner consistent with observational data. From relation \eqref{eq1:H2_small_alpha}, we can rewrite equations \eqref{30} and \eqref{30.1} in the following form,

\begin{eqnarray}
X&\approx&\Big({\rho\over3 M^2M_p^2}\Big)^{1\over\beta+1}, \\
Z&\approx&{\alpha \rho\over3M_p^2}.
\end{eqnarray}

In this regime, the system is governed by the parametric constraints $X\gg1$ and $Z\ll1$. The parameters $\alpha$, $\beta$, $n$, $M$ and $M_B$ cannot be chosen arbitrarily. To identify the physically plausible parameter regimes, we generated the contour plots shown in Fig. \ref{fig1}, focusing on the regions where $X\gg1$ and $Z\ll1$.

To investigate the sensitivity of the model to the parameters $\alpha$, $\beta$, $M$, and $M_B$, we present a series of contour plots in the $(\alpha,\beta)$ and $(M,M_B)$ planes. These visualizations identify regions that satisfy the physical constraints  $X\gg1$, $Z\ll1$, and observational constraint on $Y_B$. The top row of Figure \ref{fig1} depicts the dependence of the quantities $Z$, $X$, and $Y_B$ on the coupling parameters $\alpha$ and rainbow parameter $\beta$. As shown in the panel for $Z$, the contours exhibit a weak dependence on $\alpha$ with a slight downward trend in $\beta$, maintaining values between approximately $0.002$ and $0.01$. Conversely, the middle panel indicates that $X$ is strongly sensitive to $\beta$, increasing monotonically from $X\approx50$ to $X\approx500$ as $\beta$ increases. The behavior of $Y_B$ (right panel) demonstrates a characteristic curvature, where higher values are concentrated in the lower-right region of the $(\alpha,\beta)$ parameter space.
The bottom row of Figure \ref{fig1} illustrates the parameter dependencies in the $(M,M_B)$ plane. In the left panel, the contours of $Z$ show a distinct dependence on the mass scale $M_B$, where $Z$ increases significantly with $M_B$ while exhibiting only a slight decrease with increasing $M$. The middle panel reveals that $X$ scales positively with both mass parameters, exhibiting a diagonal gradient across the surveyed range. Finally, the $Y_B$ contours (right panel) display a downward-sloping curvature, confirming that $Y_B$ is maximized at low $M$ and high $M_B$ values. Taken together, these plots define the valid parameter space that simultaneously adheres to the model’s theoretical constraints.

The Figure \ref{fig2} illustrates the comparative behavior of the baryon-to-entropy ratio for two distinct gravitational actions within the regime $Z\ll1$. Both panels employ identical axes, with the equation of state parameter $w$ spanning the range $[-1,1/3]$ and the ratio $Y_B$ plotted up to $1.2\times10^{-9}$.
As shown in  Figure \ref{fig2} the Einstein-Hilbert contribution $Y_B^{EH}$ exhibits a parabolic, reaching a peak value of approximately $1.2\times10^{-9}$ near $w\approx-0.4$ and vanishing at $w=1/3$.
In sharp contrast, the GB contribution $Y_B^{GB}$ remains negligible throughout the majority of the specified $w$ domain, consistently maintaining values below $1\times10^{-10}$ and it becomes non-vanishing as $w$ approaches $1/3$.
These results demonstrate that the Einstein-Hilbert action provides no contribution to baryon asymmetry during the radiation-dominated epoch $(w=1/3)$. Conversely, the GB action exhibits the potential to generate a non-vanishing baryon asymmetry within this regime.
The right panel of  Figure \ref{fig2} extends this analysis to explore the sensitivity of $Y_B$ to variations in the rainbow parameter $\beta$. The results demonstrate that while the functional form of $Y_B^{EH}$ remains invariant, its peak amplitude is inversely correlated with $\beta$: increasing $\beta$ from 1.0 to 1.2 results in a systematic reduction of the peak $Y_B^{EH}$ magnitude from $1.22\times10^{-9}$ to $8.2\times10^{-10}$.

Figure \ref{fig3} illustrates the variation of the baryon-to-entropy ratio, $Y_B$, with the model parameters $\alpha$, $n$, $M$, $M_B$ and the equation-of-state parameter $w$, for three values of the rainbow parameter, $\beta\in\{1.00,1.05,1.10\}$, in the regime $Z\ll1$. The baryon asymmetry increases with $\alpha$, $n$, $M$, and $M_B$, whereas its dependence on $w$ is non-monotonic, exhibiting a peak near $w\approx-0.35$. In all panels, increasing the rainbow parameter from $\beta=1.00$ to $1.10$ systematically reduces the baryon-to-entropy ratio, while preserving the overall qualitative behavior of each curve.

To investigate the dependence of the baryon-to-entropy ratio $Y_B$ with respect to the parameters $\beta$, $n$, $M$, $w$ and $M_B$ we present Figure \ref{fig4} for various values of the rainbow parameter $\alpha$ in the regime $Z\ll1$. As shown in Figure \ref{fig4}, for a given value of the parameter $\alpha$, increasing the value of $n$, $M$ and $M_B$ increase the baryon-to-entropy ratio in the regime $Z\ll1$. $Y_B$ decreases monotonically with increasing $\beta$, indicating that larger values of the deformation parameter suppress the generated baryon asymmetry. $Y_B$ increases from zero at $w=-1$, reaches a maximum around $w\approx -0.35$, and then decreases as $w$ approaches positive values.

\begin{center}\label{tabl1}
\begin{tabular}[t]{ccccccccc}
\hline
\multicolumn{9}{c}{ Table 5.1: Evaluation of the model's parameters in $Z\ll1$ regime} \\
\hline
~$\alpha$~ &~$\beta$~ & ~$M$~ & ~$M_B$~ & ~$Z$~ & ~$X$~ &~~~~~~~$T_D$~~~~~~~ & ~~~~~$Y_B$~~& \\
\hline
$10^{15}M_p^{-2}$ & $1$ & $9\times10^{-12}{M_P}$ & $3\times10^{-4}{M_P}$ & $0.001$&$80.6$& $3.49\times10^{13}GeV$ & $5.2\times10^{-12}$& \\
\hline
$10^{16}M_p^{-2}$ & $1$ & $9\times10^{-12}{M_P}$ & $3\times10^{-4}{M_P}$ & $0.015$&$80.5$& $3.49\times10^{13}GeV$ & $5.2\times10^{-11}$& \\
\hline
$10^{17}M_p^{-2}$ & $1$ & $9\times10^{-12}{M_P}$ & $3\times10^{-4}{M_P}$ & $0.156$&$80.1$& $3.48\times10^{13}GeV$ & $5.11\times10^{-10}$& \\
\hline
$10^{16}M_p^{-2}$ & $1.1$ & $9\times10^{-12}{M_P}$ & $3\times10^{-4}{M_P}$ & $0.005$&$76.5$& $3.40\times10^{13}GeV$ & $3.8\times10^{-11}$& \\
\hline
$10^{16}M_p^{-2}$ & $1.2$ & $9\times10^{-12}{M_P}$ & $3\times10^{-4}{M_P}$ & $0.004$&$73.1$& $3.32\times10^{13}GeV$ & $2.9\times10^{-11}$& \\
\hline
$10^{16}M_p^{-2}$ & $1$ & $9\times10^{-11}{M_P}$ & $3\times10^{-4}{M_P}$ & $0.016$&$14.3$& $4.65\times10^{13}GeV$ & $1.6\times10^{-11}$& \\
\hline
\end{tabular}
\end{center}

Table \ref{tabl1} illustrates the numerical evaluation of the baryon-to-entropy ratio $Y_B$ within the $Z\ll1$ regime for radiation-dominated universe $w=1/3$. We observe that $Y_B$ is highly sensitive to the mass scales $M$ and $M_B$, as well as the gravitational coupling $\alpha$. Specifically, parameters in the third row yield $Y_B\approx5.11\times10^{-10}$, demonstrating that for a decoupling temperature $T_D\approx3.48\times10^{13} GeV$, the model successfully reproduces the observed cosmological baryon asymmetry.

\subsection{ regime $\alpha \mathcal{H}^2 \gg 1$}

In this subsection, we perform a numerical analysis of gravitational baryogenesis in the framework of 4D EGB gravity, focusing on the high-curvature regime characterized by $\alpha \mathcal{H}^2\gg1$, where the GB term provides the dominant contribution to the dynamics of the cosmological evolution. We investigate whether this regime is capable of producing a baryon-to-entropy ratio consistent with current observational constraints. From relation \eqref{HEfriedmann}, we can rewrite equations \eqref{30} and \eqref{30.1} in the following form,
\begin{eqnarray}
X&\approx&\Big({\sqrt{\rho}\over\sqrt{3\alpha} M^2M_p}\Big)^{1\over\beta+1}, \\
Z&\approx&\sqrt{\alpha \rho\over3M_p^2},
\end{eqnarray}
where in this regime $X\gg1$ and $Z\gg1$.

Figure \ref{fig5} Contour plots of $Z$, $X$, and the baryon-to-entropy ratio $Y_B$ in the $(\alpha,\beta)$ (top row) and $(M,M_B)$ (bottom row) parameter spaces, illustrating the regions compatible with the conditions $Z\gg1$ and $X\gg1$. The contours show that increasing the GB coupling $\alpha$, the mass scale $M$, and the baryon-violating scale $M_B$ generally enhances $Z$, $X$, and $Y_B$, whereas increasing the rainbow parameter $\beta$ suppresses these quantities. The largest values of the baryon-to-entropy ratio occur for large $\alpha$, $M$, and $M_B$, together with small $\beta$. The overlap between the regions satisfying $Z\gg1$ and $X\gg1$ and those yielding suitable $Y_B$ identifies the viable parameter space for successful gravitational baryogenesis within the present model.

In Figure \ref{fig6} the baryon-to-entropy ratio $Y_B$ is shown as a function of the GB coupling $\alpha$, $n$, the mass scale $M$, the baryon-violating mass scale $M_B$, and the equation-of-state parameter $w$ for three representative values of the rainbow parameter $\beta\in\{1.0,1.1,1.2\}$ in the regime $Z\gg1$. The upper-left panel shows that $Y_B$ decreases monotonically with increasing $\alpha$, while the upper-middle and upper-right panels indicate that the baryon asymmetry is enhanced by increasing $n$ and $M$, respectively. The lower-left panel reveals a non-monotonic dependence on the equation-of-state parameter, with $Y_B$ increasing from $w=-1$, reaching a maximum at $w=1/3$, and then gradually decreasing for larger $w$. The lower-right panel demonstrates that $Y_B$ increases steadily with the baryon-violating mass scale $M_B$. In all panels, increasing the rainbow parameter suppresses the generated baryon asymmetry, leading to the consistent ordering $Y_B(\beta=1.0)>Y_B(\beta=1.1)>Y_B (\beta=1.2)$, indicating that rainbow gravity effects tend to reduce the efficiency of gravitational baryogenesis in the $Z\gg1$regime.

In Figure \ref{fig7} the baryon-to-entropy ratio $Y_B$ is plotted as a function of the rainbow parameter $\beta$, $n$, the mass scale $M$, the baryon-violating mass scale $M_B$, and the equation-of-state parameter $w$ for three representative values of the GB coupling, $\alpha=10^{23} M_p^{-2}$, $5\times10^{23} M_p^{-2}$, and $10^{24} M_p^{-2}$, in the regime $Z\gg1$. The upper-left panel shows that $Y_B$ decreases rapidly with increasing $\beta$, while the upper-middle, upper-right, and lower-right panels demonstrate that the baryon asymmetry increases monotonically with $n$, $M$, and $M_B$, respectively. The lower-left panel reveals a non-monotonic dependence on the equation-of-state parameter, with $Y_B$ increasing from $w=-1$, reaching a maximum at $w=1/3$, and then gradually decreasing for larger values of $w$. In contrast to the $Z\ll1$ regime, increasing the GB coupling suppresses the generated baryon asymmetry, leading to the ordering $Y_B (\alpha=10^{23} M_p^{-2})>Y_B (\alpha=5\times10^{23} M_p^{-2})>Y_B (\alpha=10^{24} M_p^{-2})$ in all panels. These results indicate that successful gravitational baryogenesis in the $Z\gg1$ regime is favored by relatively small values of both the rainbow parameter $\beta$ and the GB coupling $\alpha$, together with sufficiently large values of $n$, $M$, and $M_B$.

\begin{center}\label{tabl2}
\begin{tabular}[t]{ccccccccc}
\hline
\multicolumn{9}{c}{ Table 5.2: Evaluation of the model's parameters in $Z\gg1$ regime} \\
\hline
~$\alpha$~ &~$\beta$~ & ~$M$~ & ~$M_B$~ & ~$Z$~ & ~$X$~ &~~~~~~~$T_D$~~~~~~~ & ~~~~~$Y_B$~~& \\
\hline
$10^{23}M_p^{-2}$ & $1$ & $10^{-12}{M_P}$ & $10^{-3}{M_P}$ & $654$&$81$& $5.9\times10^{13}GeV$ & $1.02\times10^{-10}$& \\
\hline
$10^{24}M_p^{-2}$ & $1$ & $10^{-12}{M_P}$ & $10^{-3}{M_P}$ & $1820$&$42$& $5.5\times10^{13}GeV$ & $2.20\times10^{-11}$& \\
\hline
$10^{25}M_p^{-2}$ & $1$ & $10^{-12}{M_P}$ & $10^{-3}{M_P}$ & $5066$&$22$& $5.2\times10^{13}GeV$ & $4.74\times10^{-12}$& \\
\hline
$10^{23}M_p^{-2}$ & $1.1$ & $10^{-12}{M_P}$ & $10^{-3}{M_P}$ & $624$&$64$& $5.8\times10^{13}GeV$ & $8.90\times10^{-11}$& \\
\hline
$10^{23}M_p^{-2}$ & $1.2$ & $10^{-12}{M_P}$ & $10^{-3}{M_P}$ & $599$&$52$& $5.6\times10^{13}GeV$ & $7.80\times10^{-11}$& \\
\hline
$10^{23}M_p^{-2}$ & $1$ & $10^{-11}{M_P}$ & $10^{-3}{M_P}$ & $1091$&$11$& $7.6\times10^{13}GeV$ & $4.70\times10^{-10}$& \\
\hline
$10^{23}M_p^{-2}$ & $1$ & $10^{-12}{M_P}$ & $10^{-4}{M_P}$ & $11$&$11$& $7.6\times10^{12}GeV$ & $4.70\times10^{-12}$& \\
\hline
\end{tabular}
\end{center}

The parametric sensitivity of the model is evaluated in Table \ref{tabl2}. By varying the coupling parameters $\alpha$, $\beta$ and the mass scales $M$, $M_B$, we demonstrate that the model can successfully produce the observed baryon asymmetry $(Y_B\sim10^{-10})$ within the $Z\gg1$ regime. The results indicate that a decoupling temperature $T_D\approx 10^{13}GeV$ provides a viable framework for gravitational baryogenesis under the studied parameter space.

\section{Conclusion}
We have investigated gravitational baryogenesis in regularized four-dimensional Einstein-Gauss-Bonnet (4D EGB) gravity, with Rainbow gravity corrections. The combined framework successfully resolves the radiation-era problem of standard gravitational baryogenesis by generating a non-vanishing $\dot{R}$ even for $w=\frac{1}{3}$, where the Ricci scalar vanishes in General Relativity.
The baryon-to-entropy ratio exhibits characteristic non-monotonic dependence on the equation of state parameter, peaking near
$w=-0.35$ in the perturbative regime and at $w=\frac{1}{3}$ in the high-curvature regime. The rainbow parameter,
$\beta$.  systematically suppresses the generated asymmetry, across all parameter ranges. So the quantum-gravity corrections (Rainbow Gravity) tend to reduce the efficiency of gravitational baryogenesis allowing the model to match observations without extreme fine-tuning. The decoupling temperature acquires novel scaling due to Rainbow-EGB modifications, significantly altering freeze-out dynamics compared to standard cosmology.
Comprehensive numerical analysis identifies viable parameter spaces that simultaneously satisfy theoretical constraints and reproduce the observed $Y_B\sim 10^{-10}$. Our results establish the Rainbow-EGB framework as a phenomenologically viable and theoretically consistent setting for gravitational baryogenesis, offering a natural resolution to the radiation-era problem without introducing additional fields or exotic matter. Future work may explore generalizations with both rainbow functions, dynamical equations of state, and connections to specific particle physics models.


\begin{thebibliography}{99}

\bibitem{Planck2018}
N. Aghanim et al. [Planck Collaboration],
Planck 2018 results. VI. Cosmological parameters,
Astron. Astrophys. 641, A6 (2020),
arXiv:1807.06209 [astro-ph.CO].

\bibitem{PDG2022}
R. L. Workman et al. [Particle Data Group],
Review of Particle Physics,
Prog. Theor. Exp. Phys. 2022, 083C01 (2022).

\bibitem{Sakharov1967}
A. D. Sakharov, JETP Lett. 5, 24 (1967).

\bibitem{KolbTurner}
E. W. Kolb and M. S. Turner,
\textit{The Early Universe},
Addison-Wesley, Reading, MA (1990).
\bibitem{RiottoTrodden}
A. Riotto and M. Trodden, Ann. Rev. Nucl. Part. Sci. 49, 35 (1999), arXiv:hep-ph/9901362.
\bibitem{AffleckDine}
I. Affleck and M. Dine, Nucl. Phys. B 249, 361 (1985).
\bibitem{KuzminRubakovShaposhnikov}
V. A. Kuzmin, V. A. Rubakov, and M. E. Shaposhnikov, Phys. Lett. B 155, 36 (1985).
\bibitem{FukugitaYanagida}
M. Fukugita and T. Yanagida, Phys. Lett. B 174, 45 (1986).

\bibitem{CohenKaplan}
A. G. Cohen and D. B. Kaplan, Nucl. Phys. B 308, 913 (1988).
\bibitem{spon0}
E.V. Arbuzova, A.D. Dolgov, V.A. Novikov, Phys. Rev. D 94, 123501 (2016), arXiv:1607.01247 [astro-ph.CO].
\bibitem{spon1}
M. Dubbini, O. Luongo, A. Quaranta, Phys. Rev. D 113, 043523 (2026), arXiv:2505.03644 [gr-qc].
\bibitem{Davoudiasl2004}
H. Davoudiasl, R. Kitano, G. D. Kribs, H. Murayama, and P. J. Steinhardt, Phys. Rev. Lett. 93, 201301 (2004), arXiv:hep-ph/0403019.
\bibitem{gb1}
E. Arbuzova and A. Dolgov JCAP 06(2017)001, arXiv:1702.07477 [gr-qc].
\bibitem{gb2}
 Zhong-Wen Feng et al JCAP06(2022)022, arXiv:2203.11671 [gr-qc].
 \bibitem{gb3}
M. A. Mojahed, K. Schmitz, Xun-Jie Xu, Phys. Rev. D 111, 055005 (2025), arXiv:2409.10605 [hep-ph].
\bibitem{gb4}
Kh. Saaidi, and H. Hossienkhani, Astrophys. Space Sci. 333, 305 (2011), arXiv:1010.4966 [gr-qc].
\bibitem{DeFelice2010}
A. De Felice and S. Tsujikawa, Living Rev. Relativ. 13, 3 (2010), arXiv:1002.4928 [gr-qc].
\bibitem{Lambiase2006}
G. Lambiase and S. Mohanty, Phys. Rev. D 74, 123509 (2006), arXiv:hep-ph/0504113.
\bibitem{sad1}
H. Mohseni Sadjadi, Phys. Rev. D 77, 103501 (2008), arXiv:0710.3308 [gr-qc].
\bibitem{par1}
P. Goodarzi, JHEP  02, 029 (2026), arXiv:2508.16955 [gr-qc].
\bibitem{mod1}
V.K. Oikonomou, E. N. Saridakis, Phys. Rev. D 94, 124005 (2016), arXiv:1607.08561 [gr-qc].
\bibitem{mod2}
K. Malakar, R. Mazumdar, M. M. Gohain, K. Bhuyan, arXiv:2509.03053 [gr-qc].
\bibitem{mod3}
D. F.P. Cruz, D. S. Pereira, F. S.N. Lobo, 	Nucl. Phys. B 1023 (2026) 117304, arXiv:2509.17218 [gr-qc].
\bibitem{mod4}
S. Bhattacharjee, Phys. Dark Univ. 30, 100612 (2020), arXiv:2005.05534 [gr-qc].
\bibitem{mod5}
D. S. Pereira, 	Phys. Lett. B 875, 140312 (2026), arXiv:2412.06984 [gr-qc].
\bibitem{sad2}
H. Mohseni Sadjadi, Phys. Rev. D 76, 123507 (2007), arXiv:0709.0697 [gr-qc].
\bibitem{Odintsov2016}
S. D. Odintsov and V. K. Oikonomou, Phys. Lett. B  760, 259 (2016), arXiv:1607.00545 [gr-qc].
\bibitem{Feng2008}
B. Feng, H. Li, M. Li and X. Zhang, Phys. Lett. B 620, 27 (2005), arXiv:hep-ph/0406269.
\bibitem{par2}
P. Goodarzi, Eur. Phys. J. C  83, 990 (2023), arXiv:2307.10709 [hep-th].
\bibitem{Zimdahl1996}
W. Zimdahl, Phys. Rev. D  53, 5483 (1996), arXiv:astro-ph/9601189.
\bibitem{Harko2011}
T.Harko and F.S.N.Lobo, Eur. Phys. J. C  70, 373 (2010), arXiv:1008.4193 [gr-qc].
\bibitem{LambiaseMohanty2007}
G. Lambiase and S.Mohanty, JCAP 0712, 008 (2007), arXiv:astro-ph/0611905.
\bibitem{FujiiYanagida2002}
M. Fujii, K. Hamaguchi and T.Yanagida, Phys. Rev. D 65, 115012 (2002), arXiv:hep-ph/0202210.
\bibitem{Lovelock1971}
D. Lovelock, J. Math.  Phys. 12, 498 (1971).
\bibitem{Zwiebach1985}
B. Zwiebach, Phys. Lett. B 156, 315 (1985).
\bibitem{BoulwareDeser1985}
D. G. Boulware and S. Deser, Phys. Rev. Lett. 55, 2656 (1985).
\bibitem{GlavanLin2020}
D. Glavan and C. Lin, Phys. Rev. Lett. 124, 081301 (2020), arXiv:1905.03601.
\bibitem{GB1}
R. A. Hennigar, D. Kubiznak, R. B. Mann, C. Pollack, J. High Energ. Phys. 2020, 27 (2020), arXiv:2004.09472 [gr-qc].
\bibitem{GB2}
C. M. A. Zanoletti, B. R. Hull, C. D. Leonard, R. B. Mann, JCAP 01, 043(2024), arXiv:2310.19871 [astro-ph.CO].
\bibitem{Gurses2020}
M. G\"urses, T. \c{C}. \c{S}i\c{s}man and B. Tekin, Eur. Phys. J. C 80, 647 (2020), arXiv:2004.03390 [gr-qc].
\bibitem{LuPangPope2020}
H. L\"u and Y.Pang, Phys. Lett. B 809, 135717 (2020), arXiv:2003.11552 [gr-qc].
\bibitem{Kobayashi2020}
T.Kobayashi, JCAP 07, 013 (2020), arXiv:2003.12771 [gr-qc].
\bibitem{Aoki2020}
K.~Aoki, M.~A.~Gorji and S.~Mukohyama, Phys. Lett. B 810, 135843 (2020), arXiv:2005.03859 [gr-qc].
\bibitem{MagueijoSmolin}
J. Magueijo and L. Smolin, Class. Quant. Grav. 21, 1725 (2004), arXiv:gr-qc/0305055.
\bibitem{NojiriOdintsov}
S. Nojiri and S. D. Odintsov, Int. J. Geom. Meth. Mod. Phys. 4, 115 (2007), arXiv:hep-th/0601213.
\bibitem{ata}
K. Atazadeh, Ann. Phys. 446, 169133 (2022).
\bibitem{Zwiebach}
B. Zwiebach, Phys. Lett. B 156, 315 (1985).
\bibitem{RG2}
G. Amelino-Camelia, Int. J. Mod. Phys. D 11, 35 (2002), arXiv:gr-qc/0012051.
\bibitem{LuPangPope}
H. Lu and Y. Pang and C. N. Pope, Phys. Lett. B 809, 135717 (2020), arXiv:2003.11552 [gr-qc].
\bibitem{Ling2007}
Y. Ling, JCAP 08, 017 (2007), arXiv:gr-qc/0609129.
\bibitem{Awad2013}
A. M. Awad, A. F. Ali and B. Majumder, JCAP 10, 052 (2013), arXiv:1308.4343 [gr-qc].
\bibitem{Ali2014}
A. F. Ali, Phys. Rev. D 89, 104040 (2014), arXiv:1402.5320 [hep-th].
\bibitem{Weinberg2008}
S. Weinberg, {\it Cosmology}, Oxford University Press, Oxford (2008).
\end{thebibliography}
\end{document}